\documentclass[%
 reprint,
 amsmath,amssymb,
 prl,
]{revtex4-2}
\usepackage{graphicx}
\usepackage{dcolumn}
\usepackage{color}
\usepackage{bm}
\begin{document}
\preprint{APS/123-QED}
\title{Topologically Protected Strong-Interaction of Photonics with Free Electrons}

\author{Jing Li$^{1,2}$}
\author{Yiqi Fang$^{3}$}
\author{Yunquan Liu$^{1,2,4,5}$}%
\email{Yunquan.liu@pku.edu.cn}
\affiliation{%
$^1$State Key Laboratory for Mesoscopic Physics and Collaborative Innovation Center of Quantum Matter,\\
School of Physics, Peking University, Beijing 100871, China\\
 $^2$Beijing Academy of Quantum Information Sciences, Beijing 100193, China\\
$^3$Department of Physics, Universität Konstanz,Konstanz 78464, Germany\\
$^4$Collaborative Innovation Center of Extreme Optics, Shanxi University, Taiyuan, Shanxi 030006, China\\
$^5$Peking University Yangtze Delta Institute of Optoelectronics, Nantong, Jiangsu 226010, China}

\date{\today}

\begin{abstract}
We propose a robust scheme of studying the strong interactions between free electrons and photons using topological photonics. Our study reveals that the topological corner state can be used to enhance the interaction between light and a free electron significantly. The quality factor of the topological cavity can exceed 20 000 and the corner state has a very long lifetime even after the pump pulse is off. And thus, the platform enables us to achieve a strong interaction without the need for zero delay and phase matching as in traditional photon-induced near-field electron microscopy (PINEM). This work provides the new perspective that the topological photonic structures can be utilized as a platform to shape free electron wave packets, which facilitates the control of quantum electrodynamical (QED) processes and quantum optics with free electrons in the future. 
\end{abstract}

\maketitle

The interplay between light and matter is one of the fundamental research fields in physics \cite{evansPhotonmediatedInteractionsQuantum2018,friskkockumUltrastrongCouplingLight2019,2111.07010}. One of the basic manifestations is the scattering of the relativistic free electron with a photon, from which the incident photon could transfer a portion of its energy to the recoil electron in Compton scattering. As known, a free electron and photon cannot be coupled directly in free space due to the energy-momentum mismatch. Yet, it has been shown to produce an energy spectrum with an equidistant spacing of single photon energy in the PINEM experiment, in which the electron interacts with the near field in the vicinity of a nanostructure \cite{barwickPhotoninducedNearfieldElectron2009}. Other electromagnetic media can also be employed, e.g., photonic quasiparticles (PQs) are modes satisfying Maxwell's equations in these media, such as evanescent waves on nanorods, nanotips, and mirrors \cite{riveraLightMatterInteractions2020,buhmannDispersionForcesMacroscopic2012}. Recent studies have delved into free electron interactions with photonic crystals (PhCs) and whispering-gallery modes \cite{wangCoherentInteractionFree2020,kfirControllingFreeElectrons2020}. This kind of interaction renders PINEM as a powerful tool for studying ultrafast dynamics at nanoscale \cite{polmanElectronbeamSpectroscopyNanophotonics2019,roques-carmesFreeelectronLightInteractions2023,garciadeabajoOpticalExcitationsElectron2021,liRelativisticFreeElectrons2022}. A variety of applications of PINEM have been demonstrated, e.g., time-resolved imaging, reconstruction of PhC's dispersion relation and its Bloch modes, measuring the lifetime of the mode directly becomes possible \cite{kurmanSpatiotemporalImaging2D2021a,wangCoherentInteractionFree2020}. It has also been applied to study free electron wave packet reshaping, free electron combs, and attosecond free electron pulse trains \cite{feistQuantumCoherentOptical2015,echternkampRamseytypePhaseControl2016,priebeAttosecondElectronPulse2017,vanacoreAttosecondCoherentControl2018,reinhardtTheoryShapingElectron2020,feistHighpurityFreeelectronMomentum2020,geSpatiotemporalImagingShaping2024}. More recently, PINEM has been further used to study novel phenomena in quantum optics such as entanglement between free electrons and cavity photons \cite{kfirEntanglementsElectronsCavity2019,baranesFreeElectronsCan2022,feistCavitymediatedElectronphotonPairs2022}, quantum optical excitations \cite{digiulioProbingQuantumOptical2019}, free electron qubits \cite{reinhardtFreeElectronQubits2021,tsarevFreeelectronQubitsMaximumcontrast2021}, and preparation of novel quantum states \cite{digiulioFreeelectronShapingUsing2020,hayunShapingQuantumPhotonic2021,dahanImprintingQuantumStatistics2021}.

In order to reveal the unexplored ultrastrong coupling regime, many efforts have been devoted to enhancing the coupling between free electrons and PQs \cite{dahanResonantPhasematchingLight2020,henkeIntegratedPhotonicsEnables2021,kfirControllingFreeElectrons2020}. These methods are limited by the divergence of the electron beam during the long interaction distance because of phase mismatch. If tempting to broaden the electron spectrum by increasing the laser power, it would lead to irreversible sample damage. Alternatively, the photonic flatband resonance can enable a two-order increase of the free electron Smith-Purcell radiation \cite{yangPhotonicFlatbandResonances2023}. However, the coupling between free electrons and the localized electromagnetic field remains weak. Here, we show topological photonics offers a potential solution, which yields unique, robust photonic systems to manipulate the interaction between the light field and electron. It also provides immunity to performance degradation from the fabrication imperfections and environmental variations \cite{segevTopologicalPhotonicsWhere2020,tangTopologicalPhotonicCrystals2022,lanBriefReviewTopological2022,wangShortReviewAllDielectric2022}. 

\begin{figure}
\includegraphics[width=1.0\columnwidth]{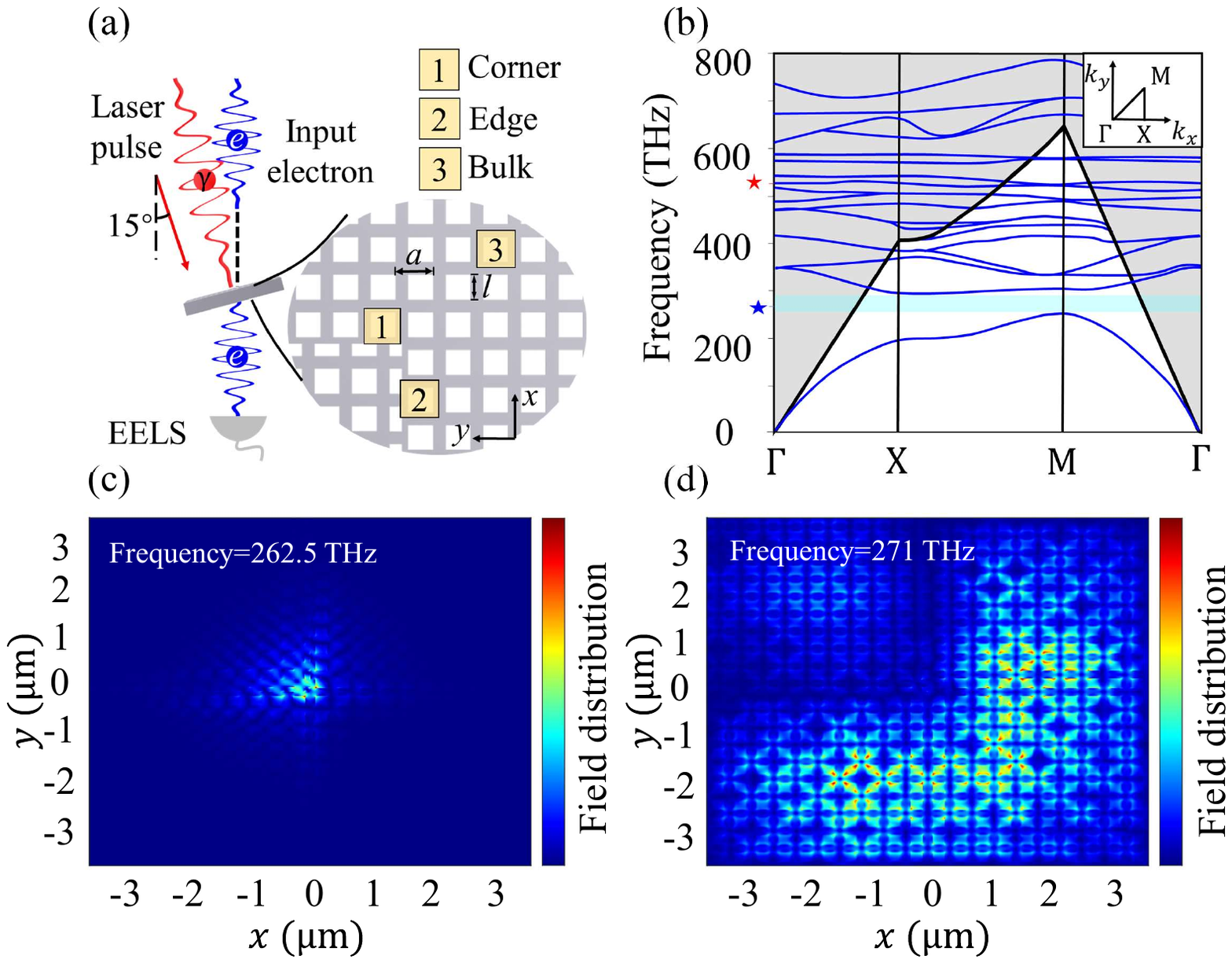}
\caption{\label{fig:1}Schematic diagram of sample. (a) The sample is tilted by 15° relative to the horizontal plane to allow for interaction between free electrons and the localized light field on the PhC slab. The inset displays the structure of the topological PhC, which is constructed by two GaAs-based PhCs with different topologies joined together. The lattice constant is $a$ and the side length of the square air hole is $l$. (b) Band structure of a PhC slab whose unit cell is indicated by any of the three yellow shadings in the inset of (a). A complete band gap appears between the two lowest bands (indicated by the light blue shading). The light cone is shown by the thick black solid line. Above the light cone, the leaky modes exist. The blue and red pentagrams represent the frequencies used to excite a corner state and a leaky mode, respectively. The inset denotes the reduced first Brillouin zone. (c) and (d) depict the field distributions of a corner state and a bulk state, respectively, with excitations at the frequencies of 271 THz and 262.5 THz. The field distributions are normalized. The band structure and field distributions are obtained by the three-dimensional finite-difference time-domain (FDTD) simulation. Perfectly matched layer boundary conditions are adopted in all the boundaries.}
\end{figure}

In this Letter, we pave a route toward the experimental realization of a robust scheme with high degrees of freedom to study the strong interactions between free electrons and topological cavity photons. The localized field enhancement effect and high quality factor of the topological corner state boost the strong interaction between PQs and the free electrons. We design a topological nanocavity with a resonance wavelength of the corner state in the near-infrared region. Thanks to the topological protection and small volume of the mode, the corner state allows for strong interaction without zero time delay and the phase-matching techniques, simplifying the experimental setup for strong-interaction-PINEM research. Secondly, we use two laser beams to excite the topological PhC. Unlike the previous work which focused the two-color laser beams on a single-crystalline graphite flake \cite{priebeAttosecondElectronPulse2017}, we find that the wave function of the free electron can be coherently shaped with the topological cavity effect, which may facilitate the control of the QED process involving free electrons \cite{wongControlQuantumElectrodynamical2021}. Notably, the interaction can be enhanced by the subsequent laser pulses at will, due to the long lifetime of the topological corner state.

The topological corner state is a kind of higher-order topological state, which has a long lifetime and highly confined field distribution \cite{otaPhotonicCrystalNanocavity2019}. These unique properties make it an ideal platform for studying light-matter interactions. It has been demonstrated that the topological corner state can enhance the light-matter interactions and has been used to generate multipolar laser modes \cite{kimMultipolarLasingModes2020}, lower the lasing threshold of nanolasers \cite{dikopoltsevTopologicalInsulatorVerticalcavity2021,yangTopologicalcavitySurfaceemittingLaser2022}, and increase the intensity and emission rate of quantum dot photoluminescence \cite{xieCavityQuantumElectrodynamics2020}. Here, we design a topological photonic nanocavity as shown in Fig. 1(a), which consists of two PhCs with different topologies. Both the PhCs have the same period of $a$ and are obtained by etching square air holes with a side length of $l=0.55\,a$ on the GaAs substrate. 
\begin{figure}
\includegraphics[width=1.0\columnwidth]{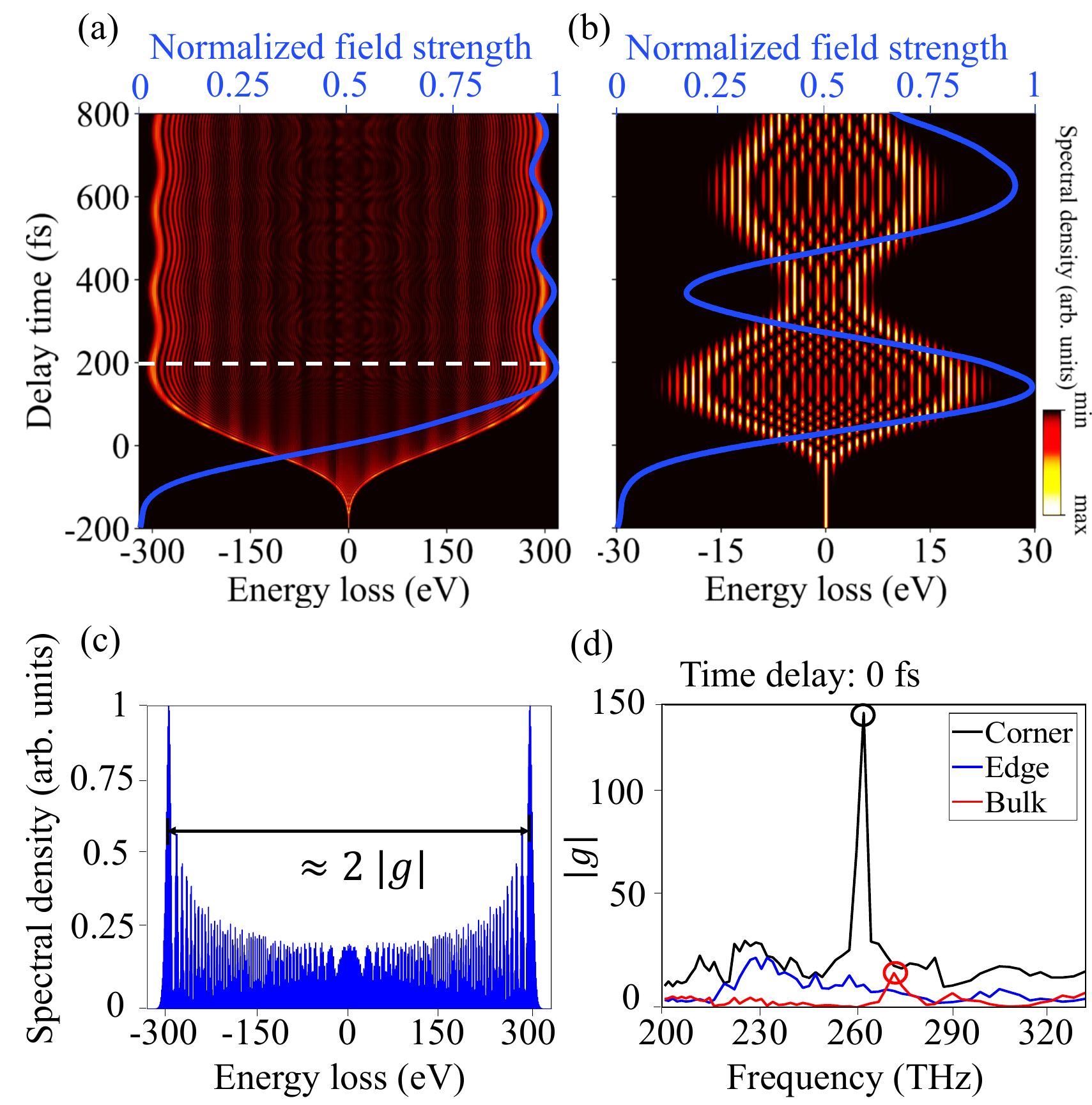}
c\caption{\label{fig:2}Free electron interacting with the topological nanocavity. (a) and (b) depict the PINEM signals of the corner state and the bulk state with the time delay between the free electron and pump laser pulses, respectively. The solid blue curve indicates the normalized field strength at the interface of PhC slab where free electrons pass through. (c) The PINEM spectrum of the corner state at 200 fs delay, indicated by the white dashed line in (a). The PINEM field strength can be roughly estimated as the half-width of EELS. (d) The modulus of the PINEM field with the varied frequency of pump laser at zero delay. The signals are measured at the corner, edge, and bulk, respectively, and the interaction regions are indicated by yellow shadings in the inset of Fig. 1(a). The black and red circles represent the corner state and bulk state used in (a) and (b), respectively.}
\end{figure}
The band structure of the PhC slab with $l=0.55\;a$ is shown in Fig. 1(b). Above the light cone [gray shading in Fig. 1(b)], the leaky modes exist. Leaky modes hold a much shorter lifetime than the bounded modes (i.e., the modes below the light cone) as a result of the coupling with the free space \cite{ochiaiDispersionRelationOptical2001a}.  A complete band gap appears between the two lowest bands, indicated by the blue shading in the figure. The corner states can be induced at the corner of the interface between the two jointed PhCs. The emergence of the topological corner state can be predicted using the bulk-edge and edge-corner correspondence principle \cite{hatsugaiChernNumberEdge1993a,trifunovicBulkandedgeCornerCorrespondence2020}. The origin of the corner states can be described by the two-dimensional Su-Schrieffer-Heeger model \cite{suSolitonsPolyacetylene1979}. The Zak phase, defined as the integral of the Berry connection of the band within the first Brillouin zone, is used to describe the topology of the band \cite{zakBerryPhaseEnergy1989}. For the topological PhC we consider, it has nonzero Zak phases in both $x$ and $y$ directions. According to the bulk-edge correspondence principle, the topological edge states will appear at one-dimensional boundaries in both $x$ and $y$ directions. These edge states can be regarded as the projections of the bulk dipole polarization dipole moment vector onto the boundary, namely, $p_x$ and $p_y$. The two dipoles form a quadrupole moment $Q_{xy}$, and the corner state is a double projection of this quadrupole tensor at the corner of the interface between the two jointed PhCs. Since the unit cell has the inversion symmetry, the edge polarization is protected by the topological property and is robust against perturbations. The corner states inherit this robustness.

The topological structure can be fabricated in a 160-nm-thick GaAs membrane using electron beam lithography and ion etching methods. The photonic-crystal sample can be then put on a TEM sample holder that allows tilting around the $x$ and $y$ axes. The pump laser is vertically focused onto the surface of the sample. The wavelength can be adjusted to excite a specified mode. We use the excitation frequency for the corner state with \~{}262.5 THz (wavelength\~{}1142 nm) and that of the bulk state with \~{}271 THz (wavelength\~{}1106 nm). The pump laser pulse duration is 100 fs and the single pulse energy is \~{}13 pJ. A motorized stage can be used to control the delay time between the electron pulse and the pump pulse. The calculated electric field distributions of a corner state and a bulk state are shown in Figs. 1(c) and 1(d), respectively.  The distribution of the corner state is highly localized, while the bulk state is diffused over the entire PhC slab. We find that the quality factor of the corner state is around 20,000, while that of the bulk state is only about 200. The corner state can be estimated with the dimensionless lifetime $Q$, which persists as long as 80 ps even the pump pulse faded off rapidly. The free electron is incident along the $-z$ axis from the midpoint of the right boundary of the air holes [indicated by regions 1 and 3 of Fig. 1(a)], and the energy of the free electron is 100 keV. Since both the excited corner state and the bulk state are TE modes (i.e., $E_z=0$), the topological PhC slab is tilted by 15° relative to the horizontal plane to allow for the interactions between free electrons and the TE modes. 

Within the single photon weak coupling limit and nonrecoil approximation \cite{garciadeabajoOpticalExcitationsElectron2021}, the interaction between the free electron and light field with a frequency of $\omega$ can be described by the PINEM field $g (t)$, where $t$ is the time delay between the laser and the electron pulse. The probability of gain or loss of $N$ photons after the interaction with a quasimonochromatic light field is given by \cite{wangCoherentInteractionFree2020}
\begin{equation}
P_N(t)=J_N^2\left(\left|g(t)\right|\right) * G\left(t, \sigma_{e}\right),
\end{equation}
with the cavity-enhanced PINEM field defined as
\begin{equation}
g(t)=\left[\left(\frac{\Theta(t) \mathrm{e}^{-\frac{t}{\tau}}}{\tau}\right) * \mathrm{e}^{-\left(\frac{t}{\sigma_l}\right)^2}\right]\frac{e}{\hbar \omega}\int_{-\infty}^{+\infty}E_z(z)\mathrm{e}^{-\mathrm{i}\Delta k z} \mathrm{d}z,
\end{equation}where $*$ represents convolution, $J_N$ is the $N$th-order Bessel function of the first kind. $e$ is the electron charge. $\Delta k$ is the electron momentum change \cite{feistQuantumCoherentOptical2015}. $G(t,\sigma_e)=\frac{1}{\sigma_e \sqrt{\pi}}\mathrm{e}^{-(t/\sigma_e)^2}$ accounts for the finite size of free electron wave packet. The term $\mathrm{e}^{-\left(t / \sigma_1\right)^2}$ describes the Gaussian distribution of the excitation laser pulse. $\Theta(t)$ is the Heaviside function, and $\tau,\, \sigma_l, \,\sigma_e$ are the cavity mode's lifetime, the standard deviation of the laser pulse, and that of the electron pulse, respectively. The detailed derivation can be found in  \cite{SM}.

By adjusting the time delay between the laser and electron pulses, we obtain the time-dependent electron energy loss spectrum (EELS) for the interaction with the corner state by performing a 3D finite-difference time-domain (FDTD) simulation with the analytical results Eqs.(1) and (2)
 from solving the time-dependent Schrödinger equation, as depicted in Fig. 2(a). Very interestingly, due to the large quality factor of the topological corner state, the strong interaction persists robustly even after a time delay of approximately 1 ps. Similarly, we calculate the EELS for the interaction with the bulk state, as illustrated in Fig. 2(b). The calculation for the edge state is given in \cite{SM}. Since the bulk state is dispersed over the entire PhC slab, the field reflects at the boundaries of the PhC slab and reexcites the PhC mode. Consequently, the EELS for the bulk state exhibits two broadened regions corresponding to the bulk state and the reexcited state. The interaction with the corner state is stronger than that with the bulk state over an order of magnitude. The strongest interaction with the corner state takes place at the time delay $\sim$ 200 fs, and the width of the energy spectrum is about 600 eV [see Fig. 2(c)]. This does not occur at the zero delay, indicating that achieving such strong interaction does not necessitate zero delay and phase matching between the electron and laser pulses. The modulus of the PINEM field, i.e., $|g|$, can be estimated as the half-width of EELS \cite{feistQuantumCoherentOptical2015}. The single-electron-single-photon coupling coefficient $g_{Qu}=g/\sqrt{\bar{n}}$ of the corner state is about 0.03, where $\bar{n}$ is the average number of photons of the pump laser pulse ($\bar{n}\approx 7.7\times 10^7$ in our cases). Hence the strong coupling with the corner state could be utilized in the electron-photon entanglement experiments \cite{feistCavitymediatedElectronphotonPairs2022,baranesFreeElectronsCan2022}.

Take it a step further, we investigate the possibility of characterizing the topological states using PINEM. We compute the responses of the $|g|$ parameter with respect to the light frequency at zero delay. For comparison, the responses at three distinct locations of the PhC [as indicated by the shallow yellow boxes in the inset of Fig. 1(a)] are computed. The results are presented in Fig. 2(d). A prominent peak at 262.5 THz is obtained at the corner, which is highlighted by a black circle. The red circle at 271 THz denotes the bulk state for comparison.

One could further distinguish between the rest bulk states and edge states by measuring time delay signals. So far, the emergence of corner state has been characterized experimentally using the microphotoluminescence spectroscopy method \cite{otaPhotonicCrystalNanocavity2019,xieCavityQuantumElectrodynamics2020}, which requires embedding quantum emitters inside the PhC slab and making measurements at low temperature. This method can only obtain signals from the $z$ component on the surface of the sample. In contrast, Eq. (2) implies that PINEM is sensitive to the buried light field of the PhC slab \cite{kurmanSpatiotemporalImaging2D2021a}. Thus PINEM may be one of the optional tools for characterizing topological photonics. 

 \begin{figure}
\includegraphics[width=1.0\columnwidth]{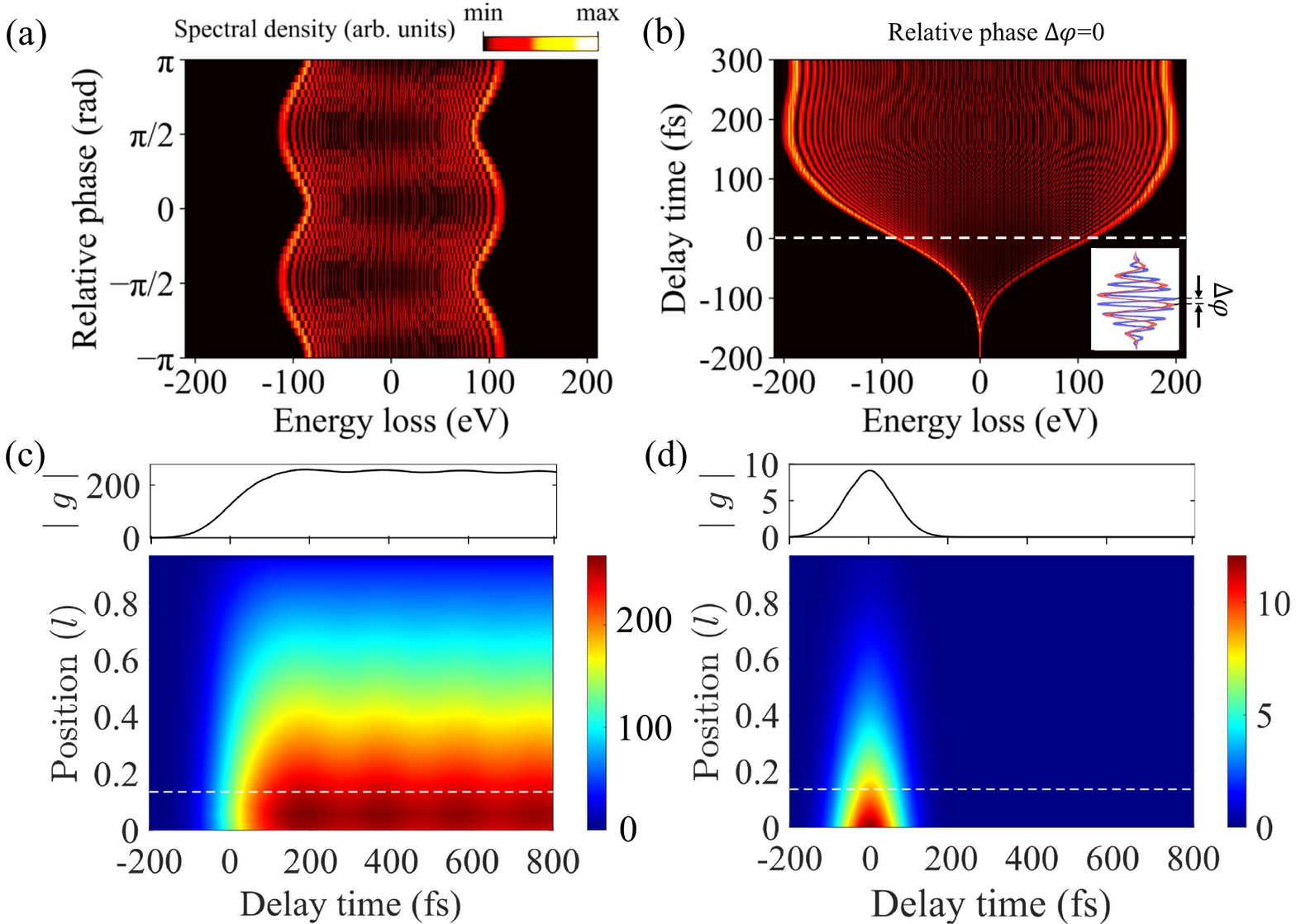}
\caption{\label{fig:3} Free electron pulse shaping using the topological PhC. (a) The EELS can be modulated by altering the relative phase between the two-color laser beams. The EELS is calculated at the zero delay between the electron and the two synthesized pump laser pulses. (b) The time delay scan of the electron energy spectrum when the relative phase equals zero. Asymmetry occurs within the $-200$ fs to 200 fs time delay. The white dashed line denotes the zero delay. The inset shows the two-color pulse, where $\Delta \varphi$ indicates the relative phase between the two pulses. (c), (d) The time delay scans of $|g|$ along the right boundary of the air hole. (c) The time delay scans of $|g|$ of the corner state. (d) The time delay scans of $|g|$ of the leaky mode. The $|g|$ decays rapidly to zero as the pump laser terminates. The upper panels outline the $|g|$ parameters at $0.15 \; l$, indicated by the white dashed line in (c) and (d), respectively.} 
\end{figure}
\begin{figure}[b]
\includegraphics[width=1.0\columnwidth]{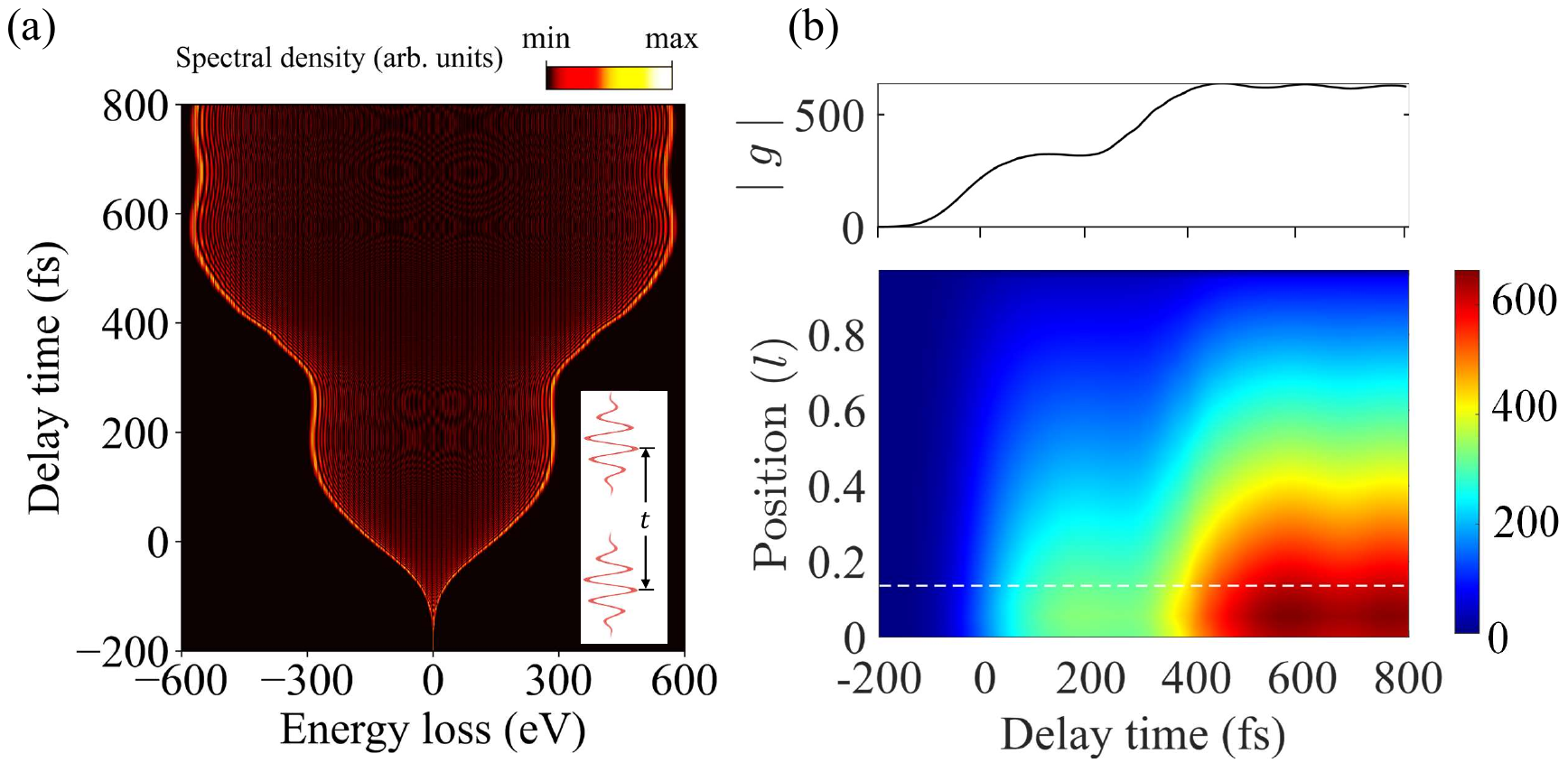}
\caption{\label{fig:4} The prolonged lifetime of the topological corner state enables one to amplify $|g|$ at will. (a) The time delay scan of the PINEM spectrum with two $\omega$ pulses in succession, resulting in the spectrum being broadened twice. The reexciting pulse of $\omega$ is delayed at 400 fs. The inset shows the two-pulse configuration. (b) The spatial distribution along the right boundary of the air hole of $|g|$ remains unchanged, but the coupling is enhanced. This implies that one could boost $|g|$ at will using the topological corner state.}
\end{figure}

To exemplify that the proposed scheme is a robust and multi-degree-of-freedom toolbox for studying light-free electron interactions, we show that one can use the topological PhC to shape the free electron wave function, thereby facilitating the control of QED processes and quantum optics with free electrons. In this instance, we employ a two-color laser field to excite the topological PhC with the help of the cavity effect. After interacting with a two-color field, the free electron wave function undergoes a nonsinusoidal phase modulation \cite{priebeAttosecondElectronPulse2017} as
\begin{equation}
\Psi(z,t)=M\left[g_\omega(t), \omega\right]\times M\left[g_{2\omega}(t), 2\omega\right]\times\psi_{\text{inc}}(z,t),
\end{equation}where we use $g_\omega (t)$ to stress that the PINEM field is dependent on the frequency of the pump laser, $M$ denotes the modulation factor. $M$ is given by
\begin{equation}
M[g_\omega(t),\omega]=\exp\left[- \mathrm{i} |g_\omega(t)| \sin\left(\frac{\omega z}{v}+\mathrm{arg}[g_\omega(t)]\right)\right],
\end{equation} where $\psi_\text{inc}(z,t)$ is the initial wave function of free electron, and $\text{arg}[g_\omega(t)]$ represents the argument of $g_\omega(t)$.
Here, we use a two-color synthesized laser field ($\omega + 2\omega$) with a 1 : 1 power ratio and scan the free electron pulses along the right boundary of the air hole.

 Adjusting the relative phase of the two pump laser beams at zero time delay, one obtains the phase-dependent EELS, as shown in Fig. 3(a). Here, the initial bandwidth of free electrons is 0.1$\,\hbar \omega$, which introduces a degree of freedom to shape the electron wave function \cite{asbanGenerationCharacterizationManipulation2021,baranesFreeElectronsCan2022}. By altering the relative phase and initial bandwidth, one could manipulate the entanglement between free electrons and photons. The relative-phase-dependent asymmetry of the PINEM spectrum occurs during the time delay of $-200$ fs to 200 fs. We have further calculated the EELS spectrum by delaying electron pulses at the relative phase equals zero [see Fig 3(b)]. When the two-color pump pulse is over, the EELS spectra become symmetrical gradually and the topologically protected effect becomes dominant. 

 One might think of the effect of the leaky mode (excited by the 2$\omega$ field) on the EELS.  We have calculated the individual contributions of the corner state and the leaky mode along the boundary of the air hole with respect to the time delay, as shown in Figs. 3(c) and 3(d). The interaction between the free electron and the leaky mode terminates when the two-color exciting pulse is off. One finds that the corner state plays the role of broadening the EELS, while the leaky mode acts as a switch, introducing asymmetry into EELS. The effect of the leaky mode will disappear when the pump pulse fades off, but the strong coupling with the corner state persists.

More intriguingly, if reexciting the corner state with another delayed $\omega$ pulse, the coupling strength can be enhanced again. As shown in Fig. 4(a), the calculated EELS spectrum with the reexciting pulse of $\omega$ at the time delay of 400 fs. One observes that the electron spectrum can be broadened twice by the second pump pulse. This would allow for enhancing the interaction at will. The spatial and time delay scans of $|g|$ are illustrated in Fig. 4(b). It's evident that the interaction is enhanced while the spatial distribution of $|g|$ remains unchanged, thus proving that the boost of $|g|$ benefits from the large quality factor and long lifetime of the corner state. Therefore, the topological corner state can also be used as the platform of the interaction of continuous wave laser with continuous wave electron beam \cite{nabbenAttosecondElectronMicroscopy2023c,tsarevFreeelectronQubitsMaximumcontrast2021}. Together with the freedom of the initial bandwidth and the degree of asymmetry, the wave function of the free electron can be shaped freely in the strong-coupling regime. Finally, one notes that, although existing works related to microresonators demonstrated higher $Q$ values than the corner state, the corner state possesses unique features, e.g., the corner state has a smaller volume of mode and topological protection. It ensures enhancement of the localized field, which is a key factor to realize strong interaction without phase matching as \cite{kfirControllingFreeElectrons2020, henkeIntegratedPhotonicsEnables2021} did. Topological protection guarantees that the cavity can be exposed to the pump laser for a long time without being overly sensitive to laser-induced damage.

\textit{Conclusion.}—We have proposed a robust high-degree-of-freedom platform for studying the strong interaction between free electrons with topological cavity photons. In this topologically protected PINEM, strong coupling does not necessitate the zero time delay and phase matching as required in traditional PINEM. It is demonstrated that higher-order topological states can be used to enhance light-matter interaction with the quality factor higher than 20 000. The topological photonic platform enables the free electron pulse shaping. Flexible controlling of the wave function of free electrons is of particular importance in controlling the QED processes. We envision that the structure can be further optimized to further enlarge the single-photon-single-electron coupling coefficient $g_{\text{Qu}}$, thus enabling the experimental demonstration of the quantum optics with free electrons \cite{sunGeneratingOpticalCat2023b}. Furthermore, one may consider the using of slow electrons for the topological photonics. When abandoning  the nonrecoil approximation \cite{talebiStrongInteractionSlow2020,talebPhaselockedPhotonElectron2023,chahshouriTailoringNearfieldmediatedPhoton2023}, it is allowed to make use of the transverse component of the corner state for the interaction between slow electrons and corner state.

\begin{acknowledgments}
 This work is supported by the National Key R$\&$D Program (No. 2022YFA1604301) and National Natural Science Foundation of China (No. 12334013, No. 92050201, and No. 92250306). We gratefully acknowledge Peter Baum, Peter Hommelhoff, Yiming Pan, Nahid Talebi, Claus Ropers, Kangpeng Wang, Jingsong Gao, and F. Javier Garcia de Abajo for illuminating and joyful discussions. 
\end{acknowledgments}

\bibliographystyle{apsrev4-2}
\bibliography{writing}

\end{document}